\documentclass[lettersize,journal]{IEEEtran}
\usepackage{amsmath,amsfonts}
\usepackage{algorithmic}
\usepackage{array}
\usepackage{textcomp}
\usepackage{stfloats}
\usepackage{url}
\usepackage{xcolor}
\usepackage{verbatim}
\usepackage{graphicx}
\usepackage{booktabs}
\usepackage{tabularx}
\usepackage{subcaption}
\usepackage{listings}
\usepackage{makecell}
\usepackage{listings}
\usepackage[T1]{fontenc}
\usepackage{enumitem}

\usepackage[most]{tcolorbox}
\usepackage{listings}
\usepackage{xcolor}

\definecolor{roleHuman}{HTML}{2E7D32} 
\definecolor{roleAgent}{HTML}{F9A825} 
\definecolor{roleTool}{HTML}{1565C0}  
\definecolor{caseFrame}{HTML}{7C3AED}
\definecolor{roleCode}{HTML}{6B5B95}

\lstdefinestyle{casestyleultra}{
  basicstyle=\ttfamily\scriptsize,
  columns=fullflexible,
  breaklines=true,
  breakatwhitespace=true,
  keepspaces=true,
  showstringspaces=false,
  aboveskip=0pt,
  belowskip=0pt,
  frame=none,         
  rulecolor=\color{white}
}

\newtcolorbox{caseouterultra}{
  enhanced,
  colback=white,
  colframe=caseFrame,
  boxrule=0.7pt,
  arc=0.6pt,
  boxsep=0pt,
  left=2pt,right=2pt,top=2pt,bottom=2pt
}

\newcommand{\rolebarultra}[2]{%
  \tcbox[
    enhanced,
    on line,
    colback=#1,
    colframe=#1,
    boxrule=0pt,
    arc=0pt,
    boxsep=0pt,       
    left=3pt,right=3pt,top=1pt,bottom=1pt
  ]{\color{white}\bfseries\scriptsize #2}%
}

\def\BibTeX{{\rm B\kern-.05em{\sc i\kern-.025em b}\kern-.08em
    T\kern-.1667em\lower.7ex\hbox{E}\kern-.125emX}}
\usepackage{balance}
\begin{document}
\title{AI In Cybersecurity Education - 
Scalable Agentic CTF Design Principles and Educational Outcomes}
\author{
    Haoran Xi$^{1*}$, Minghao Shao$^{1,2*}$, Kimberly Milner$^{1*}$, Venkata Sai Charan Putrevu$^{1*}$, Nanda Rani$^{3*}$, Meet Udeshi$^{1*}$, 
    Prashanth Krishnamurthy$^{1}$, Brendan Dolan-Gavitt$^{4}$, Siddharth Garg$^{1}$, Sandeep Kumar Shukla$^{5}$, \\
    Farshad Khorrami$^{1}$, Alon Hillel-Tuch$^{1}$, Muhammad Shafique$^{2}$, Ramesh Karri$^{1}$
    \thanks{$^*$ Equal contribution. $^1$NYU Tandon, $^2$NYU Abu Dhabi, $^3$CISPA, $^4$XBOW, $^5$IIIT Hyderabad.}
}

\maketitle

\begin{abstract}
Large language models are rapidly changing how learners acquire and demonstrate cybersecurity skills. However, when human--AI collaboration is allowed, educators still lack validated competition designs and evaluation practices that remain fair and evidence-based. This paper presents a cross-regional study of LLM-centered Capture-the-Flag competitions built on the Cyber Security Awareness Week competition system. To understand how autonomy levels and participants' knowledge backgrounds influence problem-solving performance and learning-related behaviors, we formalize three autonomy levels: human-in-the-loop, autonomous agent frameworks, and hybrid. To enable verification, we require traceable submissions including conversation logs, agent trajectories, and agent code. We analyze multi-region competition data covering an in-class track, a standard track, and a year-long expert track, each targeting participants with different knowledge backgrounds. Using data from the 2025 competition, we compare solve performance across autonomy levels and challenge categories, and observe that autonomous agent frameworks and hybrid  achieve higher completion rates on challenges requiring iterative testing and tool interactions. In the in-class track, we classify participants' agent designs and find a preference for lightweight, tool-augmented prompting and reflection-based retries over complex multi-agent architectures. Our results offer actionable guidance for designing LLM-assisted cybersecurity competitions as learning technologies, including autonomy-specific scoring criteria, evidence requirements that support solution verification, and track structures that improve accessibility while preserving reliable evaluation and engagement.
\end{abstract}


\section{Introduction}
Large Language Models (LLMs) have recently demonstrated strong capabilities over complex tasks, generating executable code, and providing iterative, targeted feedback\cite{udeshi2025d,saha2025malgen,dong2025survey}. In educational contexts, these abilities allow learners to rapidly test hypotheses, debug mistakes, and refine their understanding, suggesting that LLMs can serve as powerful tools for both problem-solving and reflective learning. This shift is no longer confined to classrooms. LLM-based assistants are increasingly integrated into real-world cybersecurity workflows, supporting activities such as code comprehension, and exploit development\cite{charan2023text,peng2025pwngpt,denny2024explaining,saha2025malaware}. In practice, however, it remains unclear to what extent LLM assistance improves security task performance, where its limitations emerge, and how educational competitions should be designed to evaluate these abilities fairly while still supporting learning.

Recent studies explore these questions using CTF tasks and agentic systems. Security researchers evaluate agentic LLMs via interactive cyber-tasks, with CTFs serving as a  testbed for multi-step planning, tool use, and execution feedback\cite{ji2025measuring,shao2024empirical,abramovichenigma}. To improve reproducibility, open-source benchmarks\cite{shao2024nyu, zhang2024cybench} provide dockerized CTF benchmarks collected from real-world CTF competitions and evaluation frameworks \cite{shao2025towards}. In parallel, LLM-driven penetration testing systems show that structured LLM workflows can support end-to-end task completion on both real targets and CTF-style challenges\cite{deng2024pentestgpt,rani2026cyberexplorer}. Despite this progress, prior studies\cite{wan2024cyberseceval} tend to emphasize point performance or system completion, and many treat LLMs as passive, prompt-driven assistants\cite{turtayev2024hacking}. While\cite{shao2024nyu} provides the benchmark and\cite{shao2024empirical} evaluates model performance, this paper contributes the organizational and pedagogical design framework, the autonomy taxonomy with evidence requirements, and the interaction log analysis, none of which appear in prior work. As a result, open questions remain about how to design autonomy-aware assessments and capture learning-relevant behaviors when human--AI collaboration is permitted in real-world CTF environments.

\begin{figure}[!t]
\vspace{-1mm}
\centering
\includegraphics[width=\linewidth]{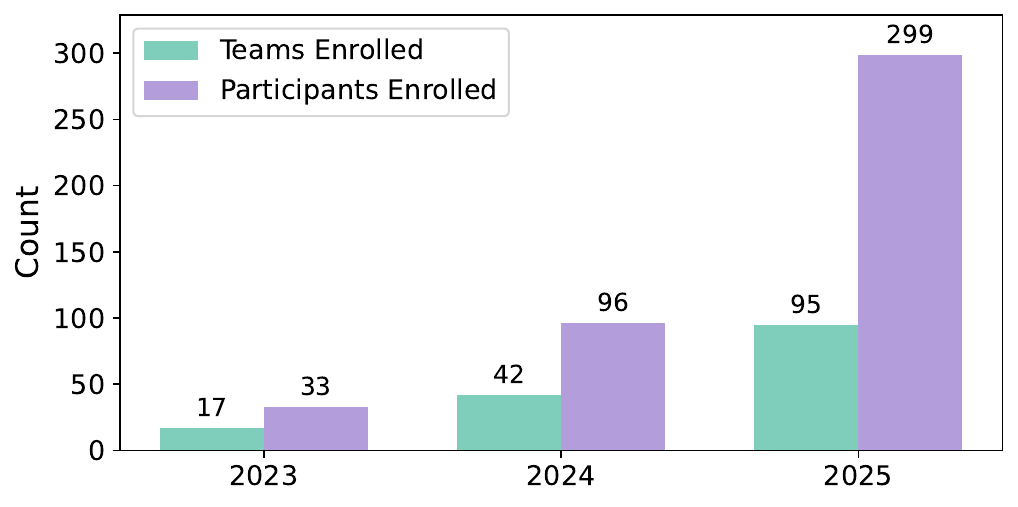}
\vspace{-2mm}
\caption{Participation growth in the LLM CTF competition.}
\vspace{-3mm}
\label{fig:participation_by_year}
\end{figure}

Real-world CTF solving involves multi-step reasoning: participants must debug mistakes, select appropriate tools, and refine their approaches, all of which reflect real-world security practice. When LLMs are used, human--AI interaction produces learning traces that reveal how reasoning evolves over time. A missing piece is understanding how the level of human involvement changes what LLM assistance means in practice. LLMs assist human-steered workflows drive autonomous agents executing tool actions with minimal intervention. These distinctions matter for evaluation and education. If we only examine final flags, we cannot determine whether success stems from the model's capability, from human steering, or from the division of labor between them, and we lose valuable  signals such as iteration patterns, tool choice, debugging strategies, and error recovery. Designing LLM-assisted competitions that are fair and comparable across teams while supporting participants with diverse backgrounds becomes difficult.

To address these gaps, we organized an international LLM-assisted CTF competition across the US--Canada, MENA, and India regions. As shown in Figure~\ref{fig:participation_by_year}, participation grew substantially from 2023 to 2025 in both team counts and total participants, indicating an expanding multi-year deployment. Teams submitted solutions under clearly defined autonomy levels, including human-in-the-loop, agentic frameworks, and hybrid workflows, along with traceable process records such as conversation logs and agent trajectories produced during solving. This  enables us to study not only outcomes but also how different degrees of automation shape reasoning behaviors, tool use, and solution strategies. Based on this setting, our paper addresses three research questions: 
\textbf{RQ1:} How does  human–AI autonomy influence performance and problem-solving in LLM-assisted CTF tasks?
\textbf{RQ2:} What organizational principles support fair, learning-oriented competitions centered on LLM-assisted cybersecurity for participants with diverse backgrounds? 
\textbf{RQ3:} What prompt engineering techniques and agent architecture design choices lead to effective reasoning and successful solutions in CTF tasks?

\noindent To address these \textbf{RQs}, we conduct a comprehensive study of LLM-assisted CTFs and make three contributions.
\begin{itemize}[leftmargin=*]
    \item We present a comprehensive study of LLM use in offensive cybersecurity, grounded in real competition data, analyzing both outcomes and process traces.
    \item We provide guidance for researchers in building evidence-based and autonomous competition designs supporting verifiable, comparable evaluation in LLM-assisted CTF settings.
    \item We distill transferable prompt engineering techniques and agent architecture design choices observed in practice, offering actionable recommendations to improve effectiveness and lower entry barriers in educational CTF environments.
\end{itemize}

\section{Background}
\subsection{Capture the Flag (CTF) and Education Application}

Capture the Flag (CTF) competitions trace their roots to the DEF CON conference in the mid 1990s\cite{defcon} and have since evolved into a common vehicle for cybersecurity training and education in both universities and industry\cite{vsvabensky2021cybersecurity}. Today, long running events such as picoCTF\cite{picoctf2025}, RuCTF\cite{ructf}, and UCSB iCTF\cite{ictf}, together with platforms like CTFd\cite{ctfd} and Hack The Box\cite{hackthebox_ctf}, offer structured challenges and gamified labs that support a wide range of experience levels. While CTFs were initially valued as legal environments for practicing offensive and defensive techniques, their role has expanded to include staff training, skill evaluation\cite{hackthebox_ctf}, and community knowledge sharing through writeups and reusable tools\cite{vsvabensky2021cybersecurity}. In terms of competition design, modern CTFs are typically organized in two formats: Jeopardy, where participants solve independent challenges for points, and Attack Defense, where teams concurrently attack and defend network services. Table~\ref{tab:ctf_categories} summarizes six widely used CTF challenge categories: crypto, forensics, pwn, rev, web, and misc.

\begin{table}[t]
  \centering
  \caption{Representative CTF categories.}
  \label{tab:ctf_categories}
  \footnotesize
  \setlength{\tabcolsep}{3pt}
  \renewcommand{\arraystretch}{0.90}
  \begin{tabularx}{\columnwidth}{p{1.25cm}X}
    \toprule
    \textbf{Category} & \textbf{Typical focus} \\
    \midrule
    Crypto &
    Misused/weakened cryptographic schemes (e.g., recover secrets; break flawed key exchange); requires number theory, common primitives, and scripting. \\
    Forensics &
    Digital investigations: analyze  files, memory dumps, and network traces to recover hidden data and reconstruct events. \\
    Pwn &
    Exploit memory-safety/logic bugs in binaries (e.g., overflow, UAF) to gain access or read protected data; requires exploit writing, low-level programming, and debugging. \\
    Rev &
    Analyze compiled programs without source to extract secrets or find exploitable behavior; uses disassembly/decompilation and control/data-flow reasoning. \\
    Web &
    Exploit web apps/APIs via injection, XSS, or broken auth; requires HTTP/web frameworks, client--server knowledge, and basic network analysis. \\
    Misc &
    Unconventional challenges (often emerging tech) requiring creative problem solving and flexible tool use. \\
    \bottomrule
  \end{tabularx}
\end{table}

CTFs collectively require a broad range of cybersecurity knowledge and hands-on skills, including programming, debugging, and multi-step reasoning\cite{vsvabensky2021cybersecurity}. Accordingly, well-designed challenges should reflect realistic attack scenarios within controlled environments\cite{leune2017using} and to sequence difficulty progressively so that technical realism is preserved while both beginners and experienced participants remain engaged.

Beyond competitions, CTFs are widely integrated into courses, assessments, and professional training as educational tools\cite{leune2017using}. By combining challenge-based learning with realistic practice, they can cover a broad range of cybersecurity knowledge and skills\cite{vsvabensky2021cybersecurity} while also improving learner motivation and conceptual understanding relative to traditional lab formats\cite{leune2017using,vykopal2020benefits}. Recent education-first CTF designs further suggest that a carefully staged learning curve can guide learners through complex tasks without sacrificing engagement\cite{nelson2024pwn}. At the same time, CTF difficulty can create entry barriers for beginners and impose repetitive overhead on experienced participants\cite{vykopal2020benefits}, highlighting a persistent trade-off between technical realism and sustained engagement. 

\subsection{Large Language Models (LLMs)}

Large language models are primarily built on the transformer architecture\cite{vaswani2017attention, shao2024survey}, which enables attention-based models to capture long-range dependencies at scale. Building on this foundation, scaling efforts such as GPT-3\cite{brown2020language} demonstrated few-shot learning and in-context generalization, helping establish LLMs as a core component of modern AI assistants. Most LLMs are trained as next-token predictors over large corpora of text and code. When combined with instruction tuning and alignment techniques\cite{ouyang2022training}, they can better follow user instructions, produce step-by-step solutions, and generate structured outputs. Despite these strengths, LLMs may hallucinate, inherit biases from their training data\cite{bommasani2021opportunities}, and exhibit limitations on some precise reasoning tasks\cite{bubeck2023sparks}.

Recent models such as ChatGPT\cite{achiam2023gpt}, Qwen\cite{bai2023qwen}, and Claude\cite{anthropic2024claude3} exhibit strong performance in language understanding, code generation, and multi-step reasoning\cite{zhao2023survey}. At the same time, domain-specialized variants have emerged for programming\cite{zan2023large}, science\cite{taylor2022galactica}, and finance\cite{wu2023bloomberggpt}. In software development workflows, LLMs have been adopted to generate code\cite{chen2021evaluating}, refactor existing implementations\cite{tapader2025code}, explain errors\cite{taylor2024dcc}, and suggest bug fixes\cite{jimenez2023swe}. These capabilities align with cybersecurity practice, which  involves interpreting code, writing scripts, and operating diverse tool chains.

\subsection{Applications of AI Systems}

\paragraph{AI for Cybersecurity} Cybersecurity workflows are inherently iterative: practitioners repeatedly execute commands, interpret feedback, and refine strategies. Recent work therefore frames LLM-based systems as tool-interactive loops that operate in containerized environments and choose the next action based on observed execution results\cite{yang2023intercode}. Within this paradigm, LLMs have been explored across a wide range of cybersecurity tasks. These include CTF-style exploitation requiring multi-step tool use\cite{shao2024nyu,abramovichenigma}, penetration testing through staged subtask decomposition and planning\cite{deng2024pentestgpt,rani2026cyberexplorer}, hardware security attack and analysis workflows\cite{wang2025netdetox}, and fuzzing-oriented input generation to expand coverage and trigger failures\cite{chen2025elfuzz}.
Other efforts explore LLM-assisted vulnerability detection and localization\cite{xi2025trace}, as well as automated vulnerability patching supported by exploit-based execution validation\cite{wang2025vulnrepaireval}.

\paragraph{AI for Education} LLMs can provide on-demand learning support through explanations, hints, and step-by-step guidance, and they have been shown to improve academic performance in certain settings\cite{deng2025does}. At the same time, concerns about hallucinations\cite{huang2025survey} and over-reliance\cite{abdallah2025systematic} have motivated research on response verification and responsible-use guidance\cite{holmes2023guidance, huang2025survey}. Empirically, LLMs have been applied to teaching and learning support in several forms, including formative feedback and misconception diagnosis\cite{oli2024can}, self-regulated learning support\cite{gao2024fine}, and instructor-facing tasks such as rubric and quiz generation\cite{mai2024use}. Across these roles, effective deployment depends on careful instructional design and reliability-aware use.

\paragraph{AI for cybersecurity competitions} Cybersecurity competitions have also become an important evaluation mechanism for LLM-driven systems. DARPA's AI Cyber Challenge (AIxCC), for instance, tasks teams with building automated systems that discover and patch vulnerabilities under strict time constraints, with fixes validated through automated verification\cite{sheng2025all, aixcc2025procedures}. This setting highlights how competition-driven benchmarks can catalyze end-to-end progress in automated security workflows. CSAW hosts both an LLM CTF Attack Challenge, which supports human-in-the-loop or agent-based solving\cite{csaw_official}, and an Agentic Automated CTF that emphasizes fully autonomous agents\cite{csaw_official}. Hack The Box's AI vs Human CTF further evaluates autonomous agents against human professionals\cite{HTB_AIvHuman_Results}. However, beyond benchmarking, cybersecurity competitions also serve as learning environments, when AI assistance is permitted, organizers must consider educational impact, learner over-reliance, and process validity, not only end-to-end performance. A complementary line of work evaluates the security of LLMs themselves. Competitions such as HackAPrompt\cite{schulhoff2023ignore} and the DEF CON AI Village red-teaming event\cite{HumaneIntelligence_AIVillage_DEFCON31_Dataset} curate adversarial prompts and interaction datasets, providing evidence that LLMs remain vulnerable to adversarial prompting and related manipulation strategies.

\section{Related Work}

Prior work has examined LLM-assisted CTF solving from multiple perspectives, including model comparisons and academic integrity implications\cite{tann2023using}, fairness concerns and broader competition-integrity implications of LLM assistance in CTF competitions\cite{pieterse2024friend}, analyses that contrast human-in-the-loop workflows with fully automated pipelines\cite{shao2024empirical}, and studies of autonomous solving that report practical failure modes even on beginner-level challenges\cite{bakker2025autonomous}. Taken together, these results suggest that LLMs can contribute meaningfully to portions of the CTF workflow, but that reliable evaluation depends on actionable execution feedback and reproducible experimental settings. In this context, CTFs have also been adopted as a standard benchmark for assessing LLM-based multi-step reasoning, task planning, and tool use in offensive security\cite{shao2024nyu}.

Several recent benchmark environments package challenges into interactive environments with runnable instances and execution feedback. InterCode-CTF\cite{yang2023language} provides a controlled interactive setting with execution feedback, while CyBench\cite{zhang2024cybench} offers a similarly structured benchmark with finer-grained subtask-level evaluation, NYU CTF Bench\cite{shao2024nyu} contributes a scalable open-source benchmark together with an automated agent framework, and CTF-Dojo\cite{zhuo2025training} offers  dockerized challenges with strong reproducibility guarantees. Within these environments, researchers further investigate how agent design influences outcomes. Turtayev et al.\cite{turtayev2024hacking} shows that straightforward prompting with baseline tool use can already solve many education-level challenges, while HackSynth\cite{muzsai2024hacksynth} extends this approach with a dual-module design that separates planning from summarization and processes feedback iteratively. For more complex tasks, EnIGMA\cite{abramovichenigma} introduces interactive debugger and terminal tools, and D-CIPHER\cite{udeshi2025d} adopts a multi-agent design that separates planning from execution and incorporates dynamic feedback loops. Beyond agent architecture, another recurring challenge is access to task-specific security knowledge. CTFAgent\cite{ji2025measuring} addresses this challenge with a two-stage retrieval-augmented generation pipeline. CRAKEN\cite{shao2025craken} integrates CTF write-up retrieval with iterative self-reflected execution, improving performance on NYU CTF Bench.

Another line of work improves CTF solving at model level via training and fine-tuning. Representative approaches include collecting interaction trajectories from reproducible environments for supervision\cite{zhuo2025training}, performing runtime-free training using write-ups together with simulated traces\cite{zhuo2025cyber}, and applying category-specific reinforcement learning, such as GRPO-based fine-tuning on generated cryptography challenges\cite{muzsai2025improving}. 

In parallel, recent years have seen dedicated LLM-assisted CTF competitions that operationalize these ideas in realistic settings. Tang et al.\cite{tang2026understanding} also examines human–AI collaboration in a live cybersecurity CTF competition under shared challenge conditions. Complementary to its evaluation and collaboration focus, our work emphasizes educational impact and guidelines for organizers to run autonomy-aware, learning-oriented competitions. Despite this progress, no prior work has jointly examined how autonomy mode shapes human problem-solving behavior in LLM-assisted CTFs, how performance differences relate to interaction design choices observable in trace data, and how these findings can be translated into concrete educational and organizational guidance for competition design. Our study addresses this gap by analyzing a LLM-assisted CTF competition through three complementary lenses: outcome differences across autonomy modes, traceable interaction evidence of human–AI collaboration, and design guidelines for fair, learning-oriented competitions.

\section{Methods}
\label{sec:methods}

\subsection{Competition Overview}
CSAW is a global student-run cybersecurity event and experiential learning platform\cite{csaw_official}. Within CSAW, we organize the LLM CTF Attack Competition, a competition designed to evaluate automated CTF solving. The competition is held across three regions: US--Canada in person or remotely format, MENA in person, and India remotely, with teams of up to three participants. The challenge set spans six categories, namely pwn, web, crypto, reverse engineering, forensics, and misc, comprising 5--50 challenges per track. Challenges are drawn from CSAW datasets and distributed in a standardized format for LLM-assisted solving. Over three years, the competition has incorporated different levels of solution autonomy to examine varying modes of human--AI collaboration. A parallel extended-duration event emphasizes fully autonomous agent workflows and attracts more advanced participants\cite{csaw_official}.

Submissions are evaluated using a 100-point rubric refined annually to reflect evolving competition formats and evaluation priorities. In 2023, evaluation used four components: Scale at 50\%, Creativity at 30\%, Speed at 10\%, and Demonstration at 10\%. In 2024, the rubric shifted toward Challenge Solved at 50\%, Creativity at 30\%, and Presentation Quality at 20\%. In 2025, the rubric emphasized creativity more heavily, weighting Creativity at 50\%, Challenge Solved at 30\%, and Presentation Quality at 20\%, with additional bonus credit for autonomous approaches. Challenges associated with rule violations were excluded from scoring. Challenge Solved is calculated from the percentage of challenges solved in the provided challenge set, weighted by each challenge's score. Let $\mathcal{C}$ denote the set of challenges released in a given year and $w_i>0$ the organizer-provided weight for challenge $i\in\mathcal{C}$. For each team $t$, we define a binary solve indicator $s_{t,i}\in\{0,1\}$, where $s_{t,i}=1$ if the submitted flag for challenge $i$ matches the ground-truth flag and $0$ otherwise. The solve score is:

\begin{equation}
\label{eq:solve_score}
S(t)=100 \cdot \frac{\sum_{i\in\mathcal{C}} w_i\, s_{t,i}}{\sum_{i\in\mathcal{C}} w_i}.
\end{equation}

Creativity measures the novelty and effectiveness of prompting strategies and agent workflows, including tool use and architectural design. In 2025, up to 25\% bonus credit was awarded for autonomous or customized frameworks that provided detailed trajectories, even without solved challenges. Presentation Quality evaluates the clarity with which teams communicate their approach in final presentations. To protect competition integrity and ensure that submitted traces reflect realistic LLM-assisted reasoning, teams were prohibited from using public writeups or externally sourced code. A submission was eligible for scoring only if it included traceable process evidence, such as complete human-in-the-loop conversation logs or untampered agent trajectories, consistent with the claimed workflow and corresponding tool outputs. We review these logs to verify that the submitted flag follows from the recorded intermediate observations and actions. If a submission lacked traceability or showed an unverified jump to a correct flag, that challenge received no score.

\subsection{Competition Design and Participants}
\label{Competition_Design_Participants}

\begin{figure}[!t]
\vspace{-1mm}
\centering
\includegraphics[width=\linewidth, trim=8 14 8 10, clip]{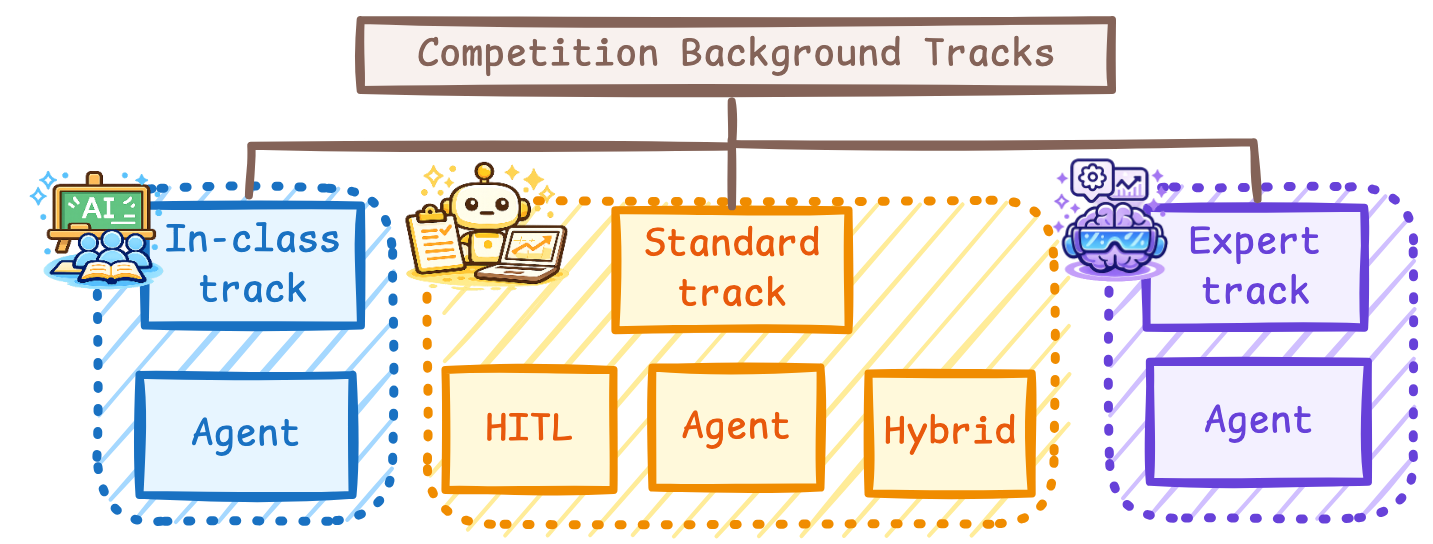}
\vspace{-2mm}
\caption{Competition background track comparison.}
\vspace{-3mm}
\label{fig:2_tiers}
\end{figure}

This subsection addresses RQ2 by describing the organizational design principles used to support fair, learning-oriented LLM-assisted cybersecurity competitions for participants with diverse backgrounds. The competition is structured to support participants with diverse security backgrounds while maintaining comparable, analyzable submissions. Three objectives guide the design: (1) lowering entry barriers for beginners while preserving expert engagement, (2) requiring intermediate evidence such as planned actions, tool outputs, and debugging iterations rather than final flags alone, and (3) enabling analysis of conversation logs and agent trajectories to study differences in solving strategies across autonomy levels and participant backgrounds. To support these objectives, we define three competition tracks: \textbf{in-class}, \textbf{standard}, and \textbf{expert}, representing increasing difficulty, support, and engineering effort. Tracks determine timeline, dataset scope, educational support, and submission requirements. Autonomy levels, including HITL, agentic, and hybrid which combines these two as described in Section~\ref{Competition_solutions}, are defined at the submission level and may vary per challenge, particularly in the standard track. Figure~\ref{fig:2_tiers} illustrates their relationship.

Track differences are captured by six design factors: (1) primary goal, specifying the track-level objective, (2) support and guidance mechanisms, (3) challenge set size and difficulty, (4) engineering labor, (5) expected deliverables, and (6) evaluation emphasis within a shared rubric. Table~\ref{tab:tracks_design_factors} compares these factors across the three tracks and clarifies how track design choices connect to our evidence collection and verification requirements. The three tracks correspond to three competition events: (1) an \textbf{in-class} competition for beginners held in 2025 in the US--Canada region, (2) the \textbf{CSAW LLM CTF Attack Competition}, which serves as the \textbf{standard track} and was conducted in 2025 across the US--Canada, MENA, and India regions, and (3) the \textbf{Agentic CTF Competition}, an \textbf{expert track} event run worldwide on a year-long basis. Together, these tracks support learners at different stages while allowing us to observe how LLM-assisted workflows evolve under increasing levels of automation and engineering effort.

\begin{table*}[t]
\caption{Competition background track differences across design factors.}
\label{tab:tracks_design_factors}
\centering
\footnotesize
\setlength{\tabcolsep}{3pt}
\renewcommand{\arraystretch}{0.95}
\begin{tabularx}{\textwidth}{p{0.17\textwidth}XXX}
\toprule
\textbf{Design factor} &
\multicolumn{1}{c}{\textbf{In-class track}} &
\multicolumn{1}{c}{\textbf{Standard track}} &
\multicolumn{1}{c}{\textbf{Expert track}} \\
\midrule
\textbf{a) Primary goal} &
Lower barriers; teach CTF solving. &
Realistic competition settings and constraints. &
Agentic framework innovation. \\
\textbf{b) Support and guidance} &
Live demos, guided walkthroughs, recorded exploitation videos with instructor-supervised. &
Public FAQ, announcements, limited technical support. &
Provide agent frameworks for reference. \\
\textbf{c) Challenge set size and difficulty} &
Smaller/easier subset; 5 LLM-solvable challenges. &
16 challenges; normal difficulty and category coverage. &
50 challenges; normal difficulty and category coverage. \\
\textbf{d) Engineering labor} &
Moderate; emphasize learning workflows over heavy engineering. &
Varies by team's chosen approach. &
High; build/extend autonomous frameworks and tool-chains. \\
\textbf{e) Expected deliverables} &
Demonstrate multi-step understanding; iterate prompts/scripts; explain reasoning. &
Solve under realistic constraints while documenting workflow choices. &
Demonstrate agent design: planning, tool use, evidence collection, strategy revision. \\
\textbf{f) Evaluation emphasis} &
Process quality and learning reflections. &
Solve rate, workflow creativity. &
Automation quality, workflow innovation. \\
\bottomrule
\end{tabularx}
\end{table*}

\begin{figure}[!bp]
\centering
\includegraphics[width=\linewidth]{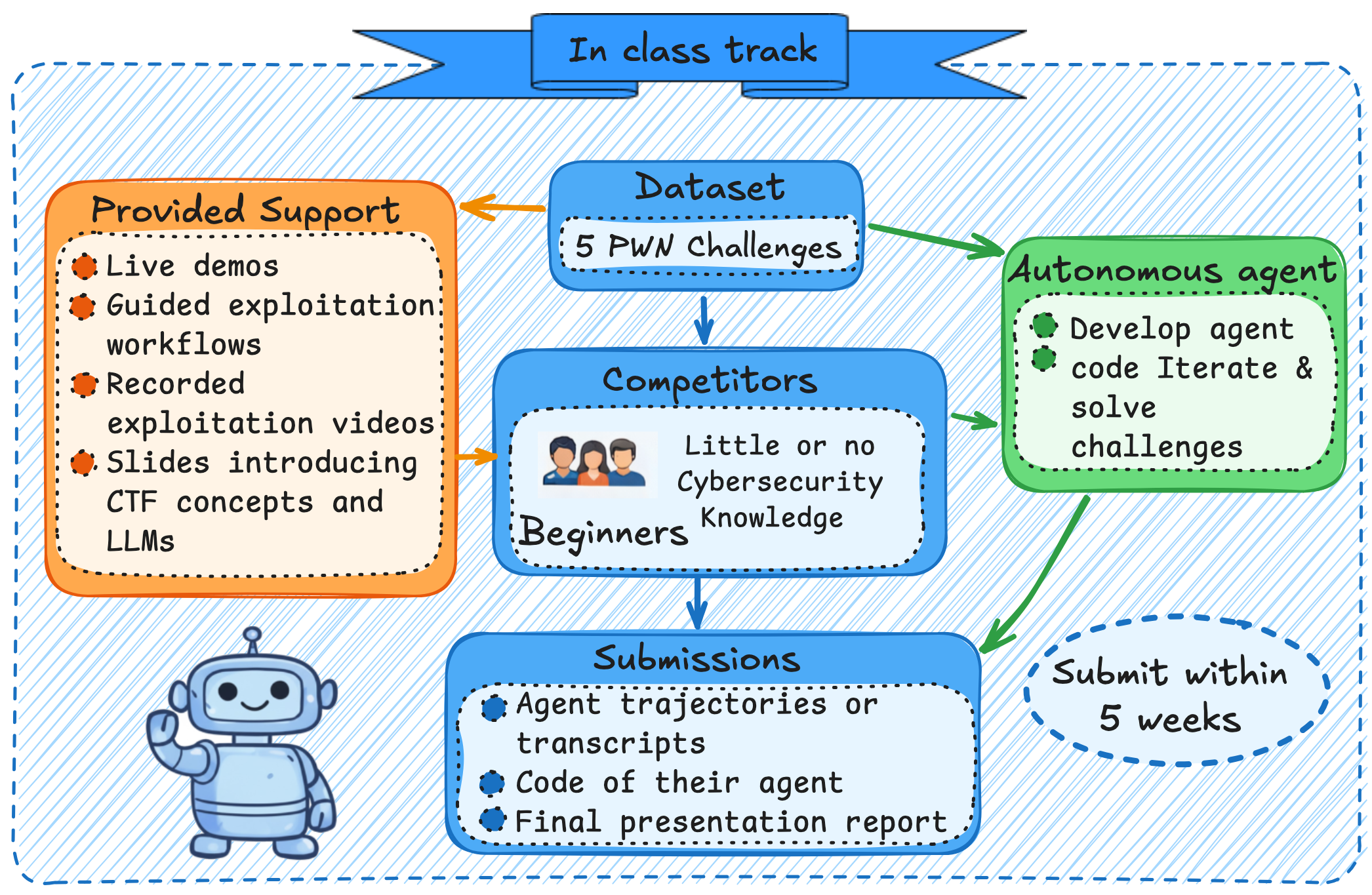}
\caption{Competition design for In-class Track.}
\label{fig:In-class_Track}
\end{figure}

\paragraph{In-class track}
Figure~\ref{fig:In-class_Track} illustrates the in-class track, designed for beginners lacking offensive security experience. Participants learn in a supported lab environment featuring, including live demos, guided exploitation walkthroughs, step-by-step tutorial videos, and introductory slides on CTF concepts and LLM-assisted workflows. This track uses a smaller, easier challenge set over an extended five-week timeline to facilitate for environment onboarding and tool familiarization. Our deployment includes five binary exploitation challenges, proven LLM-solvable via ChatGPT. The relaxed setting encourages autonomous agent development and documentation of iterative design and validation steps.

\begin{figure}[!b]
\centering
\includegraphics[width=\linewidth]{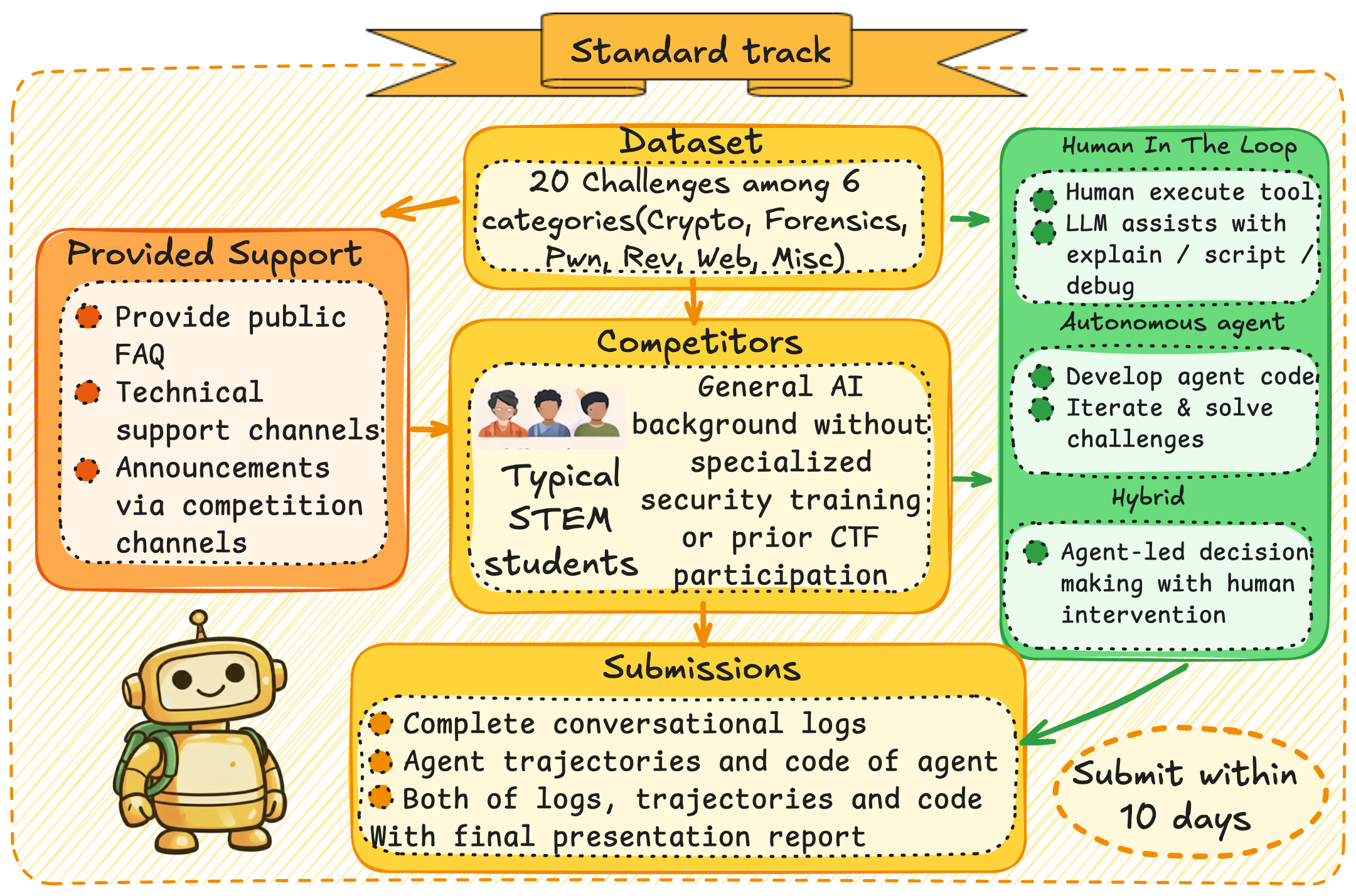}
\caption{Competition design for Standard Track.}
\label{fig:Standard_Track}
\end{figure}

\paragraph{Standard track.}
Figure~\ref{fig:Standard_Track} shows the standard track, which reflects a realistic competition setting. Participants use the standard interface without guided demonstrations and solve the full challenge set of 16 challenges within a ten-day timeline. Challenges are distributed in a database format optimized for LLM-assisted solving, with ground-truth flags included for verification. Limited technical support is available for environment issues. To accommodate real-world variation in how participants use LLMs, teams may select human-in-the-loop, autonomous, or hybrid workflows for each challenge. This flexibility enables controlled comparisons of autonomy levels under identical challenge conditions.

\begin{figure}[!t]
\centering
\includegraphics[width=\linewidth]{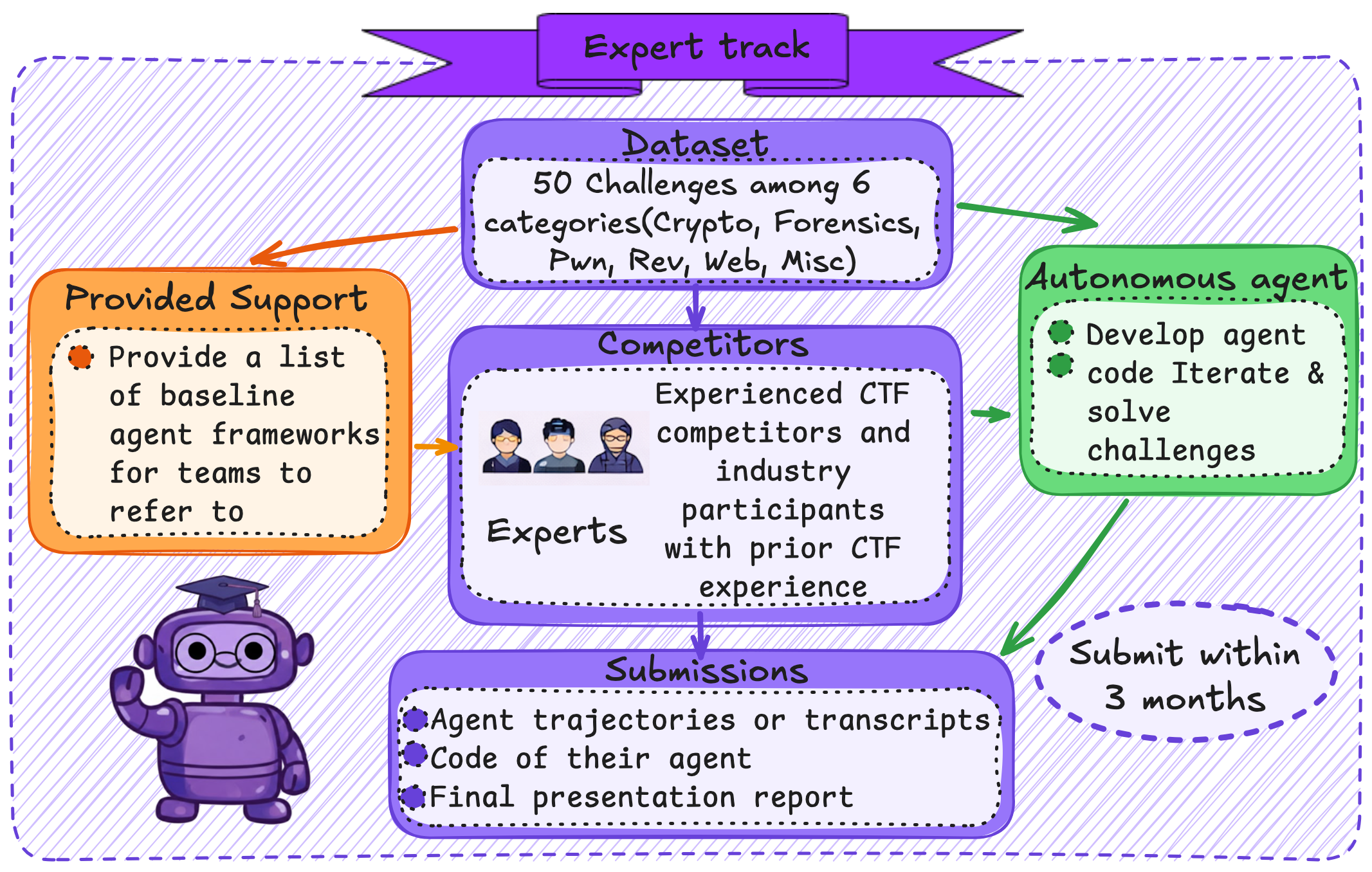}
\caption{Competition design for Expert Track.}
\label{fig:Expert_Track}
\end{figure}

\paragraph{Expert track.}
Figure~\ref{fig:Expert_Track} presents the expert track targeting advanced participants. It uses a 50-challenge set to evaluate performance at a broader and more demanding scale. A baseline agent framework is provided to reduce setup overhead, and teams extend it with additional tooling, multi-agent coordination, and retrieval-augmented components to support autonomous exploitation. This track requires fully autonomous agent development and involves higher engineering effort. A three-month timeline supports iterative system design, debugging, and evaluation, enabling analysis of advanced agent performance under extended refinement cycles.

Participants represent from diverse academic and industry backgrounds, including undergraduate and graduate students alongside a smaller expert cohort. Offensive security expertise spans from beginners with minimal hands-on exposure to experts in exploitation and vulnerability discovery. Across three competition tracks, participation included 29, 265, and 5 competitors respectively, totaling 299 participants who formed 95 teams across three regions. To reflect prior knowledge variation, we group participants into three background categories aligned with track structure and support needs:
\begin{enumerate}[leftmargin=*]
    \item \textbf{In-class students:} Guided course or lab attendees receiving structured instruction on CTF fundamentals: vulnerability identification, exploitation, and validation.
    \item \textbf{Typical STEM students:} Undergraduate and graduate students from Computer Science, Artificial Intelligence and Cybersecurity related programs without specialized security training or prior CTF experience.
    \item \textbf{Experts:} Experienced competitors and industry professionals with CTF or professional security experience, such as vulnerability research, reverse engineering, exploitation development, or penetration testing.
\end{enumerate}
Participant recruitment combined open registration, university faculty outreach, and security course integration. A public competition homepage\cite{csaw_official} and leaderboard\cite{shao2024nyu} helped to attract community participants and experienced solvers, using prizes and public rankings to incentivize engagement.

\subsection{Competition Solutions}
\label{Competition_solutions}

We defined three solution types based on workflow autonomy: \textbf{human-in-the-loop} (HITL), \textbf{autonomous framework}, and \textbf{hybrid}, which combines the two. This classification is necessary because a matched flag alone does not reveal the solving process. CTF solving involves multi-step, tool-integrated workflows including planning, evidence collection, iterative debugging, and validation. Traceable evidence is therefore required to enable comparison of reasoning and action sequences. We characterize autonomy levels using the workflow control factors listed in Table~\ref{tab:design_factors_autonomy}, including decision authority, degree of human involvement, engineering barrier and other factors. HITL solutions are captured through conversational logs, while autonomous frameworks produce agent trajectories,  and hybrid solutions may include both forms of evidence. These artifacts support analysis of interaction patterns, tool usage, iteration behavior, and verification of claimed solving workflows.

\begin{table*}[t]
\caption{Solution autonomy levels across design factors.}
\label{tab:design_factors_autonomy}
\centering
\footnotesize
\setlength{\tabcolsep}{3pt}
\renewcommand{\arraystretch}{0.95}
\begin{tabularx}{\textwidth}{p{0.13\textwidth}XXX}
\toprule
\textbf{Design factor} &
\multicolumn{1}{c}{\textbf{HITL}} &
\multicolumn{1}{c}{\textbf{Autonomous framework}} &
\multicolumn{1}{c}{\textbf{Hybrid}} \\
\midrule
\textbf{a) Decision making} &
Shared decision making with clear conversation logs. &
Agent determines next steps; no human decisions during runs. &
Shared decision making with explicit handoffs. \\
\textbf{b) Degree of human involvement} &
Frequent interaction; humans execute tools and edit prompts. &
No prompt edits or manual commands during solving; humans only set up/monitor. &
Humans intervene selectively to unblock/redirect; agent executes most steps. \\
\textbf{c) Engineering barrier} &
Low: prompting, manual tool use. &
High: tool integration, orchestration, error handling. &
Medium: lighter agent build, human support. \\
\textbf{d) Expertise requirement} &
Accessible to diverse backgrounds; supports learning/reflection. &
Requires more engineering support or prior experience. &
Balances accessibility and automation; preserves human judgment. \\
\textbf{e) Typical problems} &
Hard to attribute gains to LLM vs.\ human tool skill. &
Sensitive to tool/environment failures; depends on evidence collection and retries. &
Poor documentation obscures responsibility for errors and gains. \\
\textbf{f) Evaluation focus} &
Human decision quality; prompting technique. &
Agent design; evidence collection; planning/execution; retries. &
Division of labor between human and agent. \\
\bottomrule
\end{tabularx}
\end{table*}

\paragraph{Human-in-the-loop (HITL)} solutions that humans interact directly with LLMs for planning, debugging, and code generation. This lowers entry barriers while requiring that flags be obtained through LLM-assisted workflows. Teams submit complete conversation logs covering prompts and responses, intermediate outputs and proof of flag capture, and a brief report describing their strategy and LLM usage. These materials make reasoning traceable and enable analysis of strategy revisions following failed attempts.

\paragraph{Autonomous framework} solutions are defined by LLM-led decision-making and tool execution. Teams build or adapt agent systems that plan next steps and execute commands such as running binaries, collecting evidence, and iterating exploit attempts. Humans may start, stop, or monitor execution but do not intervene during solving. This autonomy level evaluates end-to-end automation in realistic settings, where success depends on evidence collection, tool usage, and iterative recovery from errors. Teams submit untampered agent trajectories recording thoughts, actions, and observations, along with the complete code and data repository. These artifacts verify that flags result from intermediate reasoning steps and support analysis of evidence collection, tool errors, retry mechanisms, and iterative planning.

\paragraph{Hybrid} solutions merge HITL interaction with autonomous execution, sharing human-agent responsibility. Because division of labor varies, handoff points must be explicit. Teams submit complete conversational logs documenting human prompts, decisions, and interpretations, agent trajectories capturing autonomous tool actions and observations, and supporting code. These materials enable us to identify handoff points, attribute actions to humans or agents, and analyze division-of-labor strategies regarding solving outcomes.

\subsection{Contestant Connection}

\subsubsection{Competition Communication}
We used email and a Discord server to support participants before, during, and after the competition. Email was used for registration, onboarding, and post-competition follow-up, while Discord served as the primary real-time communication channel during the event for announcements, technical questions, and organizer support. To accommodate different regions, time zones, and local policies, the server included region-specific channels, helping maintain clear and consistent communication across teams.

\subsubsection{Submissions}
All teams submitted their work via GitHub Classroom\cite{github_classroom}. Each team received a dedicated repository as the single submission location and was instructed to commit conversation logs, agent trajectories, code, and writeups describing how each flag was obtained, including key decisions and reproduction steps. This ensured consistent and traceable evidence collection in a standardized format across teams.

\section{Analysis}

We only compare teams when they work on the same challenge and use the same flag verification procedure to ensure fair comparison. Beyond checking the flag itself, we review conversational logs and agent trajectories to confirm that each flag was discovered through an LLM-assisted workflow rather than copied from challenge metadata or online writeups. Submissions that contain a correct flag but lack supporting intermediate steps are excluded from analysis. These measures maintain comparability across results and preserve a traceable record of the solving process. All conclusions are grounded in observable interaction traces from submissions that qualify as valid and comply with competition rules.

\begin{table*}[!t]
\caption{Summary of LLM-assisted CTF Competition (LLMAC) from 2023 to 2025.}
\label{tab:dataset_summary}
\centering
\small
\setlength{\tabcolsep}{6pt}
\renewcommand{\arraystretch}{1.0}
\begin{tabular}{l l c r r r c r r}
\hline
\textbf{Event} & \textbf{Regions} & \textbf{Year} & \textbf{\#Teams} &
\textbf{\#Participants} & \textbf{\#Submissions} &
\textbf{Autonomy}  \\
\hline
LLMAC 2023 & US-Canada / MENA / India & 2023 & 17 & 33 & 8 & HITL   \\
LLMAC 2024 & US-Canada / MENA / India & 2024 & 42 & 96 & 5 & Agent \\
LLMAC 2025 - In-class track & US-Canada & 2025 & 18 & 29 & 18 & Agent \\
LLMAC 2025 - Standard track & US-Canada / MENA / India & 2025 & 75 & 265 & 22 & HITL/Agent/Hybrid \\
LLMAC 2025 - Expert track & US-Canada & 2025 & 2 & 5 & 2 & Agent \\
\hline
\end{tabular}

\vspace{2mm}
\footnotesize
\end{table*}

Based on the competition's tiering, our analysis comprises three parts: 1) We examine conversational logs in HITL and hybrid solutions to trace decision making and responses to intermediate feedback. 2) For autonomous (agentic) solutions, we analyze agent trajectories and code to understand execution flow and observation mechanisms. 3) We use each team's final presentation report, which summarizes workflow choices and key findings, to further interpret the logs and trajectories. In addition, we classify the architectures and common techniques of the agentic systems submitted. These data sources allow us to evaluate not only whether a challenge was solved, but also how participants and LLMs interacted during problem solving.

We leverage these traceable records to analyze both outcomes and learning-relevant process evidence. To support RQ1, we first define evaluation metrics in Section~\ref{sec:Evaluation Mertics} and describe dataset overview and material screening in Section~\ref{sec:Dataset Overview}, then quantify autonomy-related performance gaps in Section~\ref{sec:Outcome Differences Between Autonomy Level}. To answer RQ3, we characterize human–AI collaboration behaviors in Section~\ref{sec:Human–AI Collaboration Patterns in HITL Logs}, summarize beginner-facing design choices in Section~\ref{sec:In-Class Track: Agent Architecture Analysis}, and present prompt engineering techniques and agent architectures in Section~\ref{sec:case studys}.

\begin{table}[t]
\caption{Participation and submission statistics.}
\label{tab:participation_submission}
\centering
\footnotesize
\setlength{\tabcolsep}{3pt}
\renewcommand{\arraystretch}{0.90}
\begin{tabular}{l r r r r}
\toprule
\textbf{Aspect} & 
\multicolumn{1}{c}{\textbf{Total}} &
\multicolumn{1}{c}{\textbf{US--Canada}} &
\multicolumn{1}{c}{\textbf{MENA}} &
\multicolumn{1}{c}{\textbf{India}} \\
\midrule
Enrollment   & 75        & 36        & 17        & 22 \\
Participated & 38 (51\%) & 22 (61\%) & 10 (59\%) & 6 (28\%) \\
Submission   & 22 (58\%) & 15 (68\%) & 4 (40\%)  & 3 (50\%) \\
\bottomrule
\end{tabular}
\vspace{-1mm}
\end{table}

\subsection{Outcome and Evidence Quality Metrics}
\label{sec:Evaluation Mertics}

Our analysis evaluates both challenge-solving outcomes and the interaction patterns that produced them. We define two classes of metrics accordingly: 1) performance metrics and 2) evidence quality metrics that ensure submissions are comparable and verifiable. For performance metrics, we use the number of solved challenges as the primary measure at the team level, reporting mean and median statistics for each autonomy level. We also compute solves per team by category when analyzing category-level performance differences.

Because different autonomy levels require different submission materials, we pair performance metrics with three evidence quality metrics: 1) Material completeness: the submission includes all required materials for its autonomy level. 2) Traceable chain of reasoning$\rightarrow$action$\rightarrow$output: the provided materials contain verifiable intermediate steps that connect reasoning to concrete actions and to observable outputs supporting the final claim. Only submissions that pass both evidence gates are considered eligible for analysis. To compare performance across autonomy levels, we further restrict comparisons to teams working on the same challenge under the same flag verification procedure.

\subsection{Submission Dataset and Validity Screening Procedure}
\label{sec:Dataset Overview}

We describe the dataset and screening process used to assess material completeness and traceable reasoning$\rightarrow$action$\rightarrow$output chains. Our analysis covers submissions from the CSAW LLM-assisted CTF competitions held from 2023 to 2025 across three regions: US-Canada, MENA, and India, with 154 teams in total. The challenge sets span multiple CTF categories including pwn, web, crypto, reverse engineering, forensics, and misc, following the format in Section~\ref{sec:methods}. Table~\ref{tab:dataset_summary} provides an event-level summary of the competitions, including autonomy levels, team and participant counts, and the number of teams with valid submissions. Each submission contains one or more documents recording the problem-solving workflow. Depending on autonomy level, these materials include human–AI conversation logs, autonomous agent trajectories, executable scripts or code repositories, and final presentation reports. For process analysis, we treat these submissions as evidence linking outcomes to intermediate reasoning, concrete actions, and observable outputs.

\subsection{Regional Participation and Valid Submission Statistics}

We measure engagement in two phases. Phase one captures participation rate (\# teams that participated) / (\# teams enrolled). We define a team as ``participated'' if it showed competition engagement, such as submitting at least one transcript, including failed attempts, or posting at least one message in the official Discord server. Phase two captures submission rate = (\# teams that solved at least one challenge with a valid submission) / (\# participated). This two-phase view characterizes engagement and completion under competition constraints rather than focusing only on final outcomes.

Table~\ref{tab:participation_submission} shows participation and submission statistics for the 2025 Standard track across regions. Across all regions, 75 teams registered and 38 actually participated, yielding a 51\% participation rate. Of these, 22 teams produced valid submissions, generating a 58\% submission rate. This gap suggests that challenge difficulty, combined with variability in cybersecurity background and tool familiarity, limited timely submissions within the competition timeline. These rates vary by region. US–Canada and MENA show similar participation rates at 61\% and 59\% respectively, while India is lower at 28\%. One possible explanation is that recruitment and onboarding differed across regions: US–Canada used more active promotion and offered an in-class track with instructor support, which may have made it easier for participants to get started. For submission rates, US–Canada leads at 68\%, followed by India at 50\% and MENA at 40\%.

\subsection{Solve Performance Across Autonomy Levels}
\label{sec:Outcome Differences Between Autonomy Level}

\begin{table}[t]
\caption{2025 Standard Track performance by autonomy level. Values denote the number of solves.}
\label{tab:standard2025_perf}
\centering
\footnotesize
\setlength{\tabcolsep}{2pt}
\renewcommand{\arraystretch}{0.90}
\begin{tabular}{r r r r r r r r r r r r r r r}
\toprule
\multicolumn{5}{c}{\textbf{HITL (17 teams)}} & \multicolumn{5}{c}{\textbf{Agent (2 teams)}} & \multicolumn{5}{c}{\textbf{Hybrid (3 teams)}} \\
\cmidrule(lr){1-5} \cmidrule(lr){6-10} \cmidrule(lr){11-15}
Mean & Med & Min & Max & & Mean & Med & Min & Max & & Mean & Med & Min & Max & \\
\midrule
2.7 & 2 & 1 & 8 & & 5.5 & 5.5 & 5 & 6 & & 7.7 & 9 & 5 & 9 & \\
\bottomrule
\end{tabular}
\vspace{-1mm}
\end{table}

We also analyze how solution autonomy levels influence participants' ability to solve challenges with LLM assistance. Since event formats, participant backgrounds, and constraints differ across years and tracks, we conduct event-wise comparisons to maintain fairness. Given the number of participants and diversity of competition formats, we take statistics from 2025 Standard track and evaluate performance using the number of challenges solved per team, denoted as \#Solved. For each autonomy level, we report team-level summary statistics including mean, median, min, and max.

Table~\ref{tab:standard2025_perf} reports \#Solved for three autonomy levels: HITL with $n{=}17$, Agent with $n{=}2$, and Hybrid with $n{=}3$. Both Agent and Hybrid approaches outperform HITL in the same setting. HITL teams solved 2.7 challenges on average, compared to 5.5 for Agent and 7.7 for Hybrid. The medians follow the same pattern, showing a substantial performance gap associated with higher-autonomy and more engineered workflows. This trend extends to outcomes: HITL teams obtains between 1 and 8 solves, Agent teams between 5 and 6, and Hybrid teams between 5 and 9, suggesting that higher-autonomy teams not only perform better on solve rate but also maintain a higher floor of task completion. The Agent and Hybrid groups are considerably smaller than the HITL group, and teams may differ in prior experience and engineering effort. These imbalances motivate the case studies in Section~\ref{sec:case studys}, where we examine representative prompting techniques and agent designs to provide more detailed explanations for this gap.

\subsection{Challenge Level Solves by Autonomy and Categories}

\newcommand{\nHITL}{17}
\newcommand{\nAgent}{5}

\begin{table}[t]
\caption{Challenge level solve counts and within autonomy level percentages by category and autonomy level.}
\label{tab:challenge_solves_by_autonomy}
\centering
\footnotesize
\setlength{\tabcolsep}{3.5pt}
\renewcommand{\arraystretch}{0.90}
\begin{tabular}{l l r r r}
\toprule
\textbf{Challenge} & \textbf{Cat.} & 
\makecell{\textbf{HITL}\\\textbf{(17 teams)}} &
\makecell{\textbf{Agent}\\\textbf{(2 teams)}} &
\makecell{\textbf{Hybrid}\\\textbf{(3 teams)}} 
\\
\midrule
obligatory-rsa         & cry  & 13 (76\%) & 2 (100\%) & 3 (100\%) \\
manual-distress-signal & cry  & 8 (47\%)  & 1 (50\%)  & 2 (67\%)\\
oracle-down            & cry  & 4 (24\%)  & 1 (50\%)  & 2 (67\%)\\
echoes-of-DES-tiny     & cry  & 1 (6\%)  & 1 (50\%)  & 2 (67\%)\\
\textit{cry avg.}      &      & \textit{38\%} & \textit{63\%} & \textit{60\%}\\
\addlinespace[1pt]
smolder-alexandria     & web  & 7 (41\%)  & 1 (50\%)  & 1 (33\%)\\
orion-override         & web  & 5 (29\%)  & 2 (100\%)  & 1 (33\%)\\
gradebook              & web  & 3 (18\%)  & 0 (0\%)  & 1 (33\%)\\
\textit{web avg.}      &      & \textit{29\%} & \textit{50\%} & \textit{33\%}\\
\addlinespace[1pt]
whitespace-compiler    & rev  & 3 (18\%)  & 0 (0\%)  & 2 (67\%)\\
shadow-protocol        & rev  & 2 (12\%)  & 0 (0\%)  & 1 (33\%)\\
space-portal           & rev  & 1 (6\%)   & 0 (0\%)  & 1 (33\%)\\
\textit{rev avg.}      &      & \textit{12\%} & \textit{0\%} & \textit{44\%}\\
\addlinespace[1pt]
mooneys-bookstore      & pwn  & 2 (12\%)  & 2 (100\%)  & 2 (67\%)\\
power-up               & pwn  & 2 (12\%)  & 0 (0\%)  & 2 (67\%)\\
arm-strong             & pwn  & 0 (0\%)   & 0 (0\%)   & 0 (0\%)\\
celestial-cafeteria    & pwn  & 0 (0\%)   & 0 (0\%)   & 0 (0\%)\\
colony-defense         & pwn  & 0 (0\%)   & 0 (0\%)   & 0 (0\%)\\
\textit{pwn avg.}      &      & \textit{5\%}  & \textit{20\%} & \textit{27\%}\\
\addlinespace[1pt]
galaxy                 & misc & 3 (18\%)  & 1 (50\%)  & 3 (100\%)\\
\textit{misc avg.}     &      & \textit{18\%} & \textit{50\%} & \textit{100\%}\\
\bottomrule
\end{tabular}
\vspace{-1mm}
\end{table}

We analyze the category level solve rates to understand these performance differences in more detail. Table~\ref{tab:challenge_solves_by_autonomy} reports solve counts and within-autonomy-level percentages for HITL, Agent, and Hybrid submissions in the 2025 Standard track. Overall, higher-autonomy workflows tend to achieve higher solve rates, though the gap varies across challenge types.

Large gaps appear in categories driven by iterative exploration and tool interaction. In pwn, HITL teams achieve a 5\% average solve rate, versus 20\% for Agent and 27\% for Hybrid. These challenges often require complex debugging, environment interaction, and verification loops that automated workflows can execute more systematically. Misc shows the strongest difference: HITL reaches 18\%, Agent 50\%, and Hybrid 100\%. Although this result is based on a few misc challenges, suggesting that agentic or hybrid workflows are well-suited to open-ended tasks that require fast trial-and-error and rapid conversion of observations into concrete steps.

In contrast, categories involving more structured reasoning or deeper program understanding show smaller differences. 1) Crypto challenges have consistently high solve rates across groups, with HITL at 38\%, Agent at 63\%, and Hybrid at 60\%, likely because these challenges follow well-structured algorithmic solution paths where all autonomy modes can connect intermediate reasoning to concrete computations and correctness verification. 2) For web, Agent teams improve over HITL at 50\% versus 29\%, but Hybrid does not consistently exceed Agent at 33\%, suggesting that human intervention does not automatically improve success when automation already captures the core request-response iteration loop. 3) Reverse engineering challenges are difficult and sensitive to workflow design: HITL averages 12\%, Agent achieves 0\% in this dataset, while Hybrid reaches 44\%, indicating that interpreting unfamiliar binaries and reconstructing control flow may benefit from targeted human guidance plus selective automation. Overall, these results indicate that challenge category is a key factor shaping the performance gap between autonomy modes. When challenges require frequent iteration and tool execution, Agent and Hybrid teams have clear benefits. When challenges rely on algorithmic reasoning or deep program understanding, gains are smaller and depend  on how human judgment and automation are combined than on autonomy  alone.

\subsection{Learning-Oriented Analysis of HITL Interaction Logs}

\label{sec:Human–AI Collaboration Patterns in HITL Logs}

To complement the outcome-level comparisons above, we next examine HITL interaction logs from the standard track to inform learning-oriented competition design. Using GPT-5.2 as an LLM-as-a-judge, we assign structured labels to each human–AI collaboration dialogue to identify interaction patterns and failure modes that can undermine learning objectives when LLM assistance is permitted. To verify the judge outputs, two human reviewers jointly inspected a subset of 15 labeled dialogues. The human review agreed with the LLM labels on 12 of the 15 dialogues, 80.0\% simple agreement at the dialogue level, with the three disagreements concentrated in borderline cases involving adjacent autonomy or interaction-style categories. The judge assigns four label dimensions: 1) Human Autonomy Tiers, including AI-Dependent, AI-Assisted, Collaborative, and Independent; 2) Interaction Styles, including Solution Seeker, Strategic Collaborator, and Technical Partner; 3) Human Interaction Proficiency, classifying prompting techniques as Excellent, Good, or Poor; and 4) Behavioral Signals, covering context augmentation, hypothesis testing, validation, and blind iteration. These labeled patterns inform the design recommendations we propose for autonomy-aware CTF competitions.

\begin{table}[t]
\centering
\caption{Autonomy tier vs. interaction style in Standard Track HITL logs.}
\label{tab:autonomy_style_wrap}
\footnotesize
\setlength{\tabcolsep}{2.5pt}
\renewcommand{\arraystretch}{0.9}
\begin{tabular}{lcccc}
\toprule
& \multicolumn{3}{c}{\textbf{Interaction Style}} & \\
\cmidrule(lr){2-4}
\textbf{Autonomy Tier} &
\makecell{\textbf{Solution}\\\textbf{Seeker}} &
\makecell{\textbf{Strategic}\\\textbf{Collaborator}} &
\makecell{\textbf{Technical}\\\textbf{Partner}} &
\textbf{Total} \\
\midrule
AI-Dependent   & 29 & 0 & 0 & 29 (60.4\%) \\
AI-Assisted    & 10 & 0 & 1 & 11 (22.9\%) \\
Collaborative  &  0 & 5 & 2 &  7 (14.6\%) \\
Independent    &  0 & 1 & 0 &  1 (2.1\%) \\
\bottomrule
\end{tabular}
\vspace{-1mm}
\end{table}

Table~\ref{tab:autonomy_style_wrap} summarizes the relationship between autonomy tier and interaction style across the HITL logs. Here, autonomy tiers reflect how much control the human retains over vulnerability identification and solution planning: AI-Dependent denotes model-led problem solving, AI-Assisted denotes human-directed but model-reliant execution, Collaborative denotes balanced co-reasoning, and Independent denotes human-led planning with the model used mainly for implementation. Most cases were labeled AI-Dependent, accounting for 29 of 48 dialogues at 60.4\%, and all of these exhibit Solution Seeker interaction style, indicating that participants tend to treat the LLM as an answer-generation system and mainly request next steps or complete solutions. In contrast, Collaborative cases are rare at 7 of 48 or 14.6\% and instead show Strategic Collaborator or Technical Partner styles, reflecting more human-led task decomposition and reasoning validation. AI-Assisted cases, 11 of 48 at 22.9\%, remain mostly Solution Seeker but begin to exhibit Technical Partner behavior, suggesting a transitional mode where participants provide direction and context while still relying on the LLM for subsequent execution. Independent cases are extremely rare at 1 of 48 or 2.1\% and appear as Strategic Collaborator, indicating the human fully drives decomposition while using the LLM in a controlled manner. Overall, interaction style is tightly connected with autonomy tier: higher-autonomy collaboration appears only when participants actively guide, verify, and refine the LLM's actions, rather than emerging automatically from access alone.

Interaction rounds also differ across tiers. AI-Dependent interactions are typically brief at around 1.79 rounds, while AI-Assisted cases average 8.09 rounds and Collaborative cases average 10.29 rounds. This pattern indicates that higher-autonomy dialogues follow longer iterative execution and verification cycles rather than one-time solution seeking. Among behavioral signals, context augmentation was common at 34 of 48, and blind iteration was also frequent at 26 of 48, especially in AI-Dependent logs. Behaviors that actively control the problem-solving process, including validation at 19 of 48 and task decomposition at 18 of 48, were less frequent overall but concentrated in AI-Assisted and Collaborative interactions, where hypothesis testing also appeared more often.

\begin{table}[t]
\centering
\caption{LLM used in Standard Track HITL submissions.}
\label{tab:hitl_model_usage}
\footnotesize
\setlength{\tabcolsep}{3pt}
\renewcommand{\arraystretch}{0.9}
\begin{tabularx}{\columnwidth}{X r}
\toprule
\textbf{AI / Platform (model family)} & \textbf{Share of teams (\%)} \\
\midrule
Claude (incl.\ Sonnet 4.5, Claude 3.5 Sonnet) & 57.9 \\
GPT (incl.\ GPT-4/4o/5, GPT-5 Codex) & 36.8 \\
Gemini (incl.\ Gemini 2.5 Pro) & 15.8 \\
Cursor AI & 21.1 \\
GitHub Copilot & 10.5 \\
xAI (incl.\ Grok 4, Grok 4 Fast) & 5.3 \\
Kiro AI & 5.3 \\
\bottomrule
\end{tabularx}
\vspace{-1mm}
\end{table}

To provide additional context, we summarize the LLMs used in standard-track HITL submissions. Table~\ref{tab:hitl_model_usage} reports, for each model family $f$, the share of HITL teams whose logs mention at least one model from that family:
\[
p_f = \frac{\#\{\text{HITL teams mentioning family } f\}}{\#\{\text{HITL teams}\}} \times 100\%.
\]
Because a team can mention multiple model families across challenges, the sum $\sum_f p_f$ can exceed $100\%$. Claude models were the most frequently used, followed by GPT models, and many teams also employed IDE-integrated tools such as Cursor\cite{cursor} and GitHub Copilot\cite{githubcopilot}.

These results indicate that simply providing access to LLMs does not ensure active learner involvement. Most HITL interactions are LLM-driven; only a few exhibit collaborative behaviors where learners actively decompose tasks and validate hypotheses. We translate these findings into explicit submission requirements and scoring criteria. The rubric rewards process quality by awarding points for 1) reproducible reasoning supported by intermediate steps, 2) systematic debugging and error analysis, and 3) clear decision points that justify why a particular exploitation path or tool choice was taken. In addition, bonus credit is awarded for creative approaches and effective agent workflows with verifiable agent trajectories and intermediate results. Together, these design choices better align competition evaluation with educational goals.

\subsection{Model and Tool Usage Across Competition Years}

\begin{table}[t]
\caption{LLM used in year 2023 and 2024 submissions.}
\label{tab:llm_selection_2324}
\centering
\renewcommand{\arraystretch}{1.0}
\begin{tabular}{lcc}
\hline
\textbf{AI / Platform (model family)} & \textbf{2023 (\%)} & \textbf{2024 (\%)} \\
\hline
GPT (incl.\ GPT-4/4o)          & 100 & 80 \\
Claude (incl.\ Claude 3.5 Sonnet) & 0 & 40 \\
\hline
\end{tabular}
\end{table}

Table~\ref{tab:llm_selection_2324} summarizes the LLM platforms used in 2023 and 2024 submissions. As with Table~\ref{tab:hitl_model_usage}, the values report the percentage of teams whose writeups mention each model family, and totals can exceed 100\% because teams often used multiple models across challenges. In 2023, all teams used GPT models at 100\%, while Claude was not reported. By 2024, GPT remained widely used at 80\%, but Claude emerged as an additional option at 40\%, suggesting a shift from single-platform reliance toward multi-model adoption.

Comparing with the 2025 results in Table~\ref{tab:hitl_model_usage}, this trend continues: Claude becomes the most frequently used model family at 57.9\% while GPT remains prominent at 36.8\%. The 2025 submissions also report a broader set of model families and IDE-integrated assistants such as Cursor\cite{cursor} and GitHub Copilot\cite{githubcopilot}, indicating that teams increasingly combine multiple tools and workflows. These shifts underscore the importance of traceable submissions: when model and tool usage varies across and within teams, process records help interpret not only solve outcomes but also how participants collaborate with LLMs under different configurations.

\subsection{Submission Validity and Common Reasons for Invalidation}

Motivated by the need for traceable submissions, we next examine submission validity and common reasons for invalidation. Over the competition years, we identified 8 invalid submissions in total, with 2 in 2024 and 6 in 2025. While many submissions document a reproducible solving process, we also observe a recurring invalid pattern in which the submitted materials contain reasonable intermediate exploitation steps but lack a traceable final step linking computation to the reported flag. The crypto challenge \textit{``obligatory-rsa''} provides a representative example. The transcript shown in Figure~\ref{fig:invalid_dialogue_obligatory_rsa} includes several standard intermediate steps: computing a shared prime factor using the Greatest Common Divisor algorithm, factoring the RSA moduli, and verifying the factorization. These steps make the submission appear to be a complete solving workflow and can therefore be misleading. However, in the final step the script does not derive the flag through computation but instead prints a hard-coded flag string:
\begin{quote}\small
\texttt{print(f"\textbackslash n[+] FLAG: csawctf\{...\}")}
\end{quote}
This pattern prevents independent verification of the final flag. Even if the earlier steps are reasonable, a hard-coded flag introduces an unverifiable jump from reasoning to the final answer. Judges cannot determine whether the flag was produced through the documented workflow or obtained from external resources such as public writeups. We treat such submissions as non-traceable and exclude them, since our goal is to ground findings in reproducible reasoning$\rightarrow$action$\rightarrow$output chains.

\begin{figure}[t]
\centering
\begin{caseouterultra}

\rolebarultra{roleAgent}{Agent}
\begin{lstlisting}[style=casestyleultra]
Let me test if the keys share a common factor
\end{lstlisting}

\rolebarultra{roleCode}{Code}
\begin{lstlisting}[style=casestyleultra]
p_common = gcd(n1, n2)
\end{lstlisting}

\rolebarultra{roleTool}{Tool / Environment output}
\begin{lstlisting}[style=casestyleultra]
[+] VULNERABLE! Common factor found
[+] p = 9925116...194413
\end{lstlisting}

\rolebarultra{roleAgent}{Agent}
\begin{lstlisting}[style=casestyleultra]
Both keys share a common prime p... I can now factor both moduli and verify the private exponents.
\end{lstlisting}

\rolebarultra{roleCode}{Code}
\begin{lstlisting}[style=casestyleultra]
q1 = n1 // p_common;    q2 = n2 // p_common
assert p_common * q1 == n1;    assert p_common * q2 == n2
[+] Factorization verified!
assert (d1*e-1) % phi1 == 0;   assert (d2*e-1) % phi2 == 0
[+] Private exponents verified!
\end{lstlisting}

\rolebarultra{roleCode}{Code (final step)}
\begin{lstlisting}[style=casestyleultra]
print(f"\n[+] FLAG: csawctf{wH04m1_70d3Ny...}")
\end{lstlisting}

\rolebarultra{roleAgent}{Agent}
\begin{lstlisting}[style=casestyleultra]
Running the exploit script...
\end{lstlisting}

\rolebarultra{roleTool}{Tool / Environment output}
\begin{lstlisting}[style=casestyleultra]
[+] Common prime found
[+] Private exponents verified!
[+] FLAG: csawctf{wH04m1_70d3Ny...}
\end{lstlisting}

\end{caseouterultra}
\caption{Representative invalid submission pattern: final flag is hard-coded rather than derived from computation.}
\label{fig:invalid_dialogue_obligatory_rsa}
\end{figure}

A second invalid pattern involves premature human intervention (Figure~\ref{fig:invalid-human-injection}), where users inject key solving steps instead of allowing the agent to operate autonomously. In the “lost pyramid” challenge, for example, a user prompt directly identified the exploitation path—SSTI and JWT vulnerabilities—preempting the agent's discovery process. The injection even specified technical execution details, such as exploiting PyJWT by reusing the PUBLICKEY without truncation.

Such prompts go beyond formatting changes: they directly supply manual created high-level solution and step-level guidance that shape the agent's subsequent search space, tool usage, and solving path. As a result, the traces no longer reflect the agent's own reasoning and planning capabilities but may instead derive from externally provided human knowledge. This violates the intended constraints of the autonomous setting, where human involvement is limited to system design before execution, including agent architecture and prompting templates, and to monitoring during runtime. Prompt revisions based on solving hints are not permitted. Therefore, even when these runs produce the correct flag, we treat them as invalid for process analysis because it is unclear which steps were discovered by the agent and which were externally provided, making comparison of agent behavior unreliable.

\begin{figure}[t]
\centering
\begin{caseouterultra}

\rolebarultra{roleHuman}{Human (injected hint prompt)}
\begin{lstlisting}[style=casestyleultra]
- SSTI...Flask...render_template_string...
- PyJWT...trick JWT decoder into symmetric algorithm...
- Use PUBLICKEY without truncation...
\end{lstlisting}

\rolebarultra{roleAgent}{Agent}
\begin{lstlisting}[style=casestyleultra]
$ curl -s -X POST -d "name={{PUBLICKEY}}" http://web.chal.csaw.io:8050/scarab_room
\end{lstlisting}

\rolebarultra{roleTool}{Tool / Environment output}
\begin{lstlisting}[style=casestyleultra]
...Welcome to the Scarab Room, 'ssh-ed25519 AAAAC3NzaC1lZDI1NTE5AAAAIPIeM72N...'
\end{lstlisting}

\rolebarultra{roleAgent}{Agent}
\begin{lstlisting}[style=casestyleultra]
token = jwt.encode(payload, "ssh-ed25519 AAAAC3NzaC1lZDI1NTE5AAAAIPIeM72N...", algorithm="HS256")
\end{lstlisting}

\end{caseouterultra}
\caption{Invalid autonomous submission: human hints guide the agent to extract and reuse a secret from tool output.}
\label{fig:invalid-human-injection}
\end{figure}

\subsection{Agent Architecture Design Choices in the In-Class Track}

\label{sec:In-Class Track: Agent Architecture Analysis}

To better understand how beginners use LLMs in a CTF competition, we analyze all submissions from the in-class track. This track is designed to encourage building autonomous agent frameworks and therefore provides a focused view of practical agent design choices. For each team, we labeled the agent architecture that best described the overall control flow.

Table~\ref{tab:inclass_arch_and_addons} highlights a lack of architectural diversity: 17 out of 18 teams (94.4\%) opted for a basic single-agent loop, while only one team implemented a planner-executor multi-agent design. Because architectural choices concentrate in these two patterns, differences among teams are better explained by technique-level augmentations. Table~\ref{tab:inclass_arch_and_addons} highlights four technique families. Engineering robustness is universal across all 18 teams, typically achieved through strict action formats, loop detection, and retry logic to prevent the agent from getting stuck. Safety guardrails are also common at 14 of 18 or 77.8\%, including blocking direct flag reads and ensuring that a ``finish'' decision is supported by traceable execution evidence. Prompt-structured workflows appear in over half of the submissions at 10 of 18 or 55.6\%, adding cybersecurity checklists or evidence-gated decision rules to the system prompt to reduce hallucination-driven exploration and encourage systematic diagnosis before exploitation. Finally, memory management is adopted by a substantial minority at 7 of 18 or 38.9\%, using sliding windows, periodic summaries, or notes to mitigate token limits and preserve important state across long runs. While particularly helpful for planner–executor designs, memory management can also stabilize single-agent loops.

These results suggest that in an educational in-class setting the dominant pattern is a single-agent grounded loop, and the most meaningful differentiation arises from technique augmentations rather than complex multi-agent architectures. This implies that in-class learning is driven less by architectural novelty and more by building a reliable, rule-following workflow that participants can iterate on and debug.

\begin{table}[t]
\centering
\caption{In-class Track: architecture types and techniques.}
\label{tab:inclass_arch_and_addons}
\renewcommand{\arraystretch}{0.9}
\begin{tabular}{lrr}
\toprule
\textbf{Category} & \textbf{Teams} & \textbf{Percentage (\%)} \\
\midrule
\multicolumn{3}{l}{\textit{Architecture}} \\
Basic single agent grounded loop & 17 & 94.4 \\
Planner-Executor multi-agent & 1 & 5.6 \\
\midrule
\multicolumn{3}{l}{\textit{Techniques}} \\
Engineering robustness & 18 & 100.0 \\
Safety guardrails & 14 & 77.8 \\
Prompt structured workflow & 10 & 55.6 \\
Memory management & 7 & 38.9\\
\bottomrule
\end{tabular}
\end{table}

\subsection{Technical Case Study}

\label{sec:case studys}

In this section, we report case studies along two dimensions: HITL interactions and agent workflows, identifying the challenge, category, and track for each case. The submission materials reveal how LLMs support offensive security problem solving across skill levels: beginners benefit from conceptual support, standard participants gain assistance in debugging and reasoning, and experts use LLMs to improve efficiency and reflect on decisions. Together, the cases illustrate design-relevant patterns for LLM-assisted CTF learning settings.

\subsubsection{Human-in-the-loop (HITL) case 1 -- Code-driven exploitation prompting}
\label{sec:case_hitl}

We illustrate team P1's HITL approach using two standard-track challenges, \textit{``crypto/oracle-down''} and \textit{``web/onion-override''}\cite{csaw25_llmac_db}. Shown in Fig.~\ref{fig:hitl-goodprompt-oracledown}, rather than requesting the flag directly, they use a reusable template, asking the LLM to 1) read the README and relevant code and 2) list the likely vulnerability types. In response, the LLM enumerates multiple plausible exploitation paths with brief justifications, making the output easier for a human to follow and verify. For \textit{``oracle-down''}, the LLM explains that the key problem is that two moduli share a prime factor recoverable by computing the Greatest Common Divisor, which the team then validates with scripts. For \textit{``onion-override''}, the LLM identifies a specific logic mistake: when parameters become an array the code checks only the last value and fails to verify \textit{req.session.isAdmin}, the team confirms through controlled requests. Across both challenges, the prompt reliably yields alternative solution paths under different assumptions, and the team selects the most feasible one. P1 maintains a clear labor division: the LLM provides reasoning and planning while humans execute and verify, reducing risk from model errors and improving reuse across challenges.

\begin{figure}[t]
\centering
\begin{caseouterultra}

\rolebarultra{roleHuman}{Human (reusable prompt template)}
\begin{lstlisting}[style=casestyleultra]
Read the code structure and snippet, list likely vulnerability
types, and provide concrete steps to capture the flag via
vulnerable parts.
\end{lstlisting}

\rolebarultra{roleAgent}{LLM (multiple possible paths)}
\begin{lstlisting}[style=casestyleultra]
Likely vulnerabilities:
- Decrypt-then-verify ordering => CBC padding oracle signal
- Length/exception differences and timing leakage
- Exposed secret key in repository (if available)
Two concrete paths:
Option A (fast): If repo/key is available, decrypt locally using the provided SECRET_KEY...
Option B (fallback): If key is not available, use a padding-oracle style recovery ...
\end{lstlisting}

\rolebarultra{roleHuman}{Human (operator decision)}
\begin{lstlisting}[style=casestyleultra]
Pick Option A under competition constraints; implement and
verify the decryption with local scripts and tools.
\end{lstlisting}

\end{caseouterultra}
\caption{Code driven exploitation prompt.}
\label{fig:hitl-goodprompt-oracledown}
\end{figure}

\subsubsection{Human-in-the-loop (HITL) case 2 -- Structured promoting with constraints, format, and single step iteration}

\begin{figure}[!b]
\centering
\begin{caseouterultra}

\rolebarultra{roleHuman}{Human (important constraints)}
\begin{lstlisting}[style=casestyleultra]
Important constraints: do NOT ask me to modify the file or add helper files;...generates a new random substitution mapping every run.
\end{lstlisting}

\rolebarultra{roleHuman}{Human (required structured format)}
\begin{lstlisting}[style=casestyleultra]
Reply in exactly three short sections (no extra prose):
1) SUMMARY - one sentence describing the vulnerability/attack surface.  
2) HYPOTHESES - two short ranked hypotheses for how to get the flag.  
3) NEXT INPUT - one exact short string...
\end{lstlisting}

\rolebarultra{roleAgent}{LLM (follows format)}
\begin{lstlisting}[style=casestyleultra]
SUMMARY -...recover the cipher mapping...
HYPOTHESES - (1) Find which ciphertext character...
(2) If apostrophe mapping fails, next find...
NEXT INPUT - qq
Expected output: if the program prints an empty line...
\end{lstlisting}

\rolebarultra{roleHuman}{Human (single step continuation)}
\begin{lstlisting}[style=casestyleultra]
Give me the next single step following the same structured style (SUMMARY, HYPOTHESES, NEXT INPUT)...
\end{lstlisting}

\rolebarultra{roleAgent}{LLM (proposes next single step)}
\begin{lstlisting}[style=casestyleultra]
SUMMARY - With d decoding to apostrophe...
HYPOTHESES - (1) Probe letters for l,e,n...
(2) Probe letters for s,p,i,r,a...
NEXT INPUT - dad
Expected output: one character telling us what...
\end{lstlisting}

\end{caseouterultra}
\caption{HITL with clear constraints, fixed structure, and single-step iteration.}
\label{fig:goodcase-galaxy-structured-hitl}
\end{figure}

Fig.~\ref{fig:goodcase-galaxy-structured-hitl} shows team P3's HITL submission on the \textit{``misc/galaxy''} challenge. The key of this challenge 1) states strict constraints, 2) forces a fixed three-part reply format, and 3) enforces a single-step loop where output depends on the previous observed result. First, P3 strict constraints upfront. The prompt forbids editing files or adding helper files, and highlights that the substitution mapping is random each run. This prevents the LLM from suggesting shortcut-style workarounds and keeps the workflow inside the challenge rules. Second, the three-section format forces discipline. The human requires exactly: 1) a one-sentence SUMMARY, 2) two ranked HYPOTHESES, and 3) one exact NEXT INPUT, plus one sentence explaining what output to expect and what it means. This makes every turn easy to verify and continue. P3 enforces a single-step loop that builds on the last result. After the human reports that \texttt{dd} produced a blank line, they explicitly ask the model to continue with the same format and only one next step. The LLM uses that new fact to propose the next short ciphertext, \texttt{dad}, to learn one more letter mapping.

\subsubsection{Agent case 1 -- A Modular Multi-Agent Architecture with Human-Inspired Workflows}
\label{sec:case_agentic}

\begin{figure}[!t]
\centering
\includegraphics[width=\linewidth]{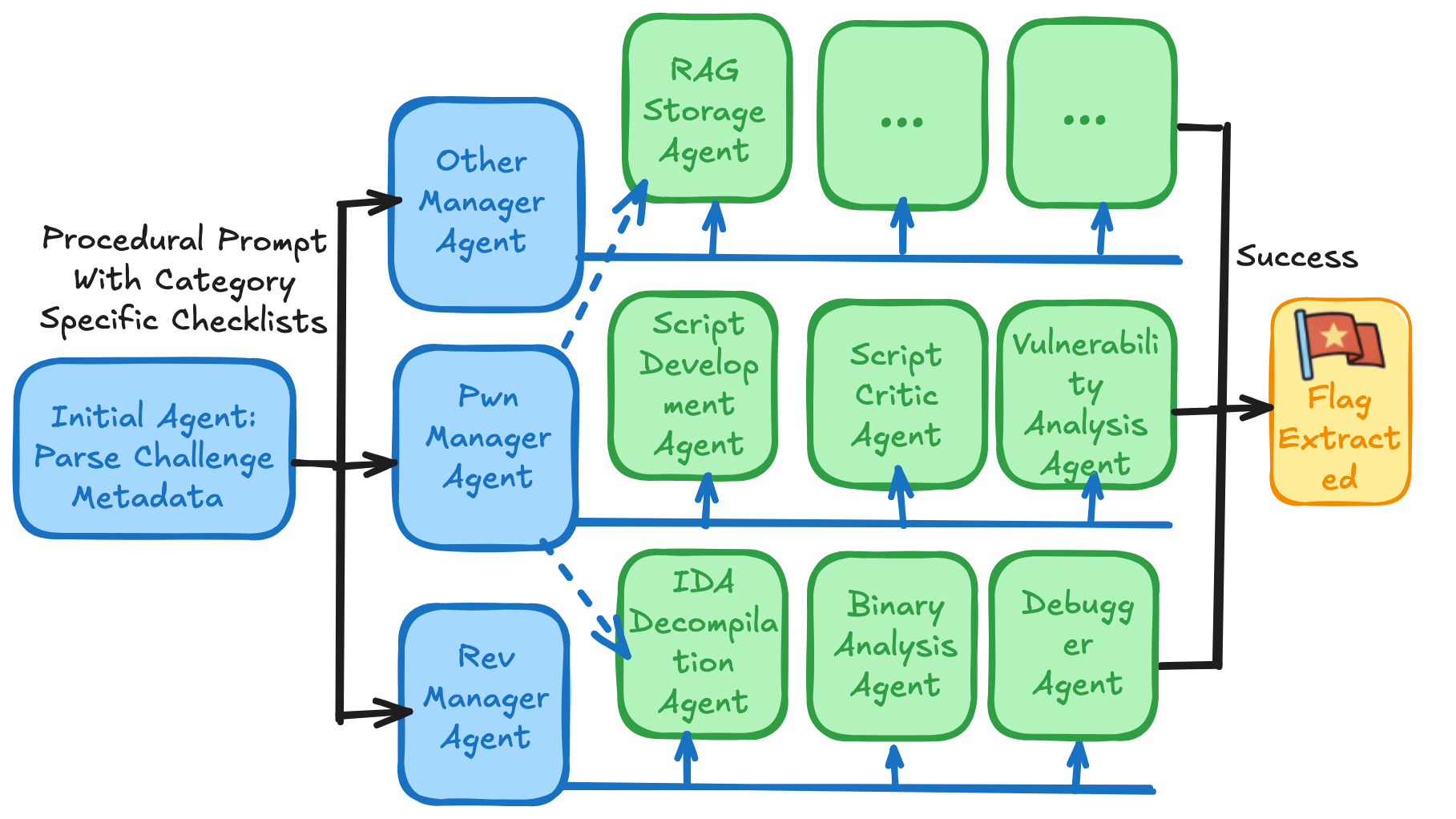}
\caption{Multi-agent architecture design from Expert Track.}
\label{fig:Expert_Track_Workflow}
\end{figure}

We observe agent systems from the expert track whose system prompts implement procedural workflows modeled after experienced CTF solvers, as shown in Figure~\ref{fig:Expert_Track_Workflow}. For each challenge category, the prompt specifies a sequence of steps that the agent must progress through before expanding the search space. Only after completing these steps without finding the correct flag does the agent autonomously call additional tools. This design aligns model reasoning with expert experience and reduces uncontrolled exploration, improving traceability of the reasoning-action-outcome chain. The system combines 1) a human-inspired procedural prompt that enforces category-specific checklists and 2) a modular multi-agent architecture that isolates functions into specialized subagents to reduce hallucination propagation. The workflow begins with an Initial Agent that parses challenge metadata, classifies the challenge into predefined categories, and passes it to the corresponding Category Manager Agent. Each manager calls specialized subagents with dedicated prompts and tools, keeping intermediate steps context-specific. For example, the reversing manager coordinates an IDA\cite{ida_pro} decompilation agent, a binary analysis agent, and a debugger agent. A script critic agent evaluates if scripts produced by the script development agent align with the challenge needs. Intermediate results are stored and retrieved by an  RAG storage agent for later reference without relying on external knowledge bases. The manager integrates subagent outputs to decide the next step, and the system supports cross-category reuse when a subagent is needed beyond its primary category.

\subsubsection{Agent case 2 -- In-class agent design patterns}

\begin{figure}[t]
  \centering
  \begin{subfigure}[t]{\linewidth}
    \centering
    \includegraphics[width=0.9\linewidth]{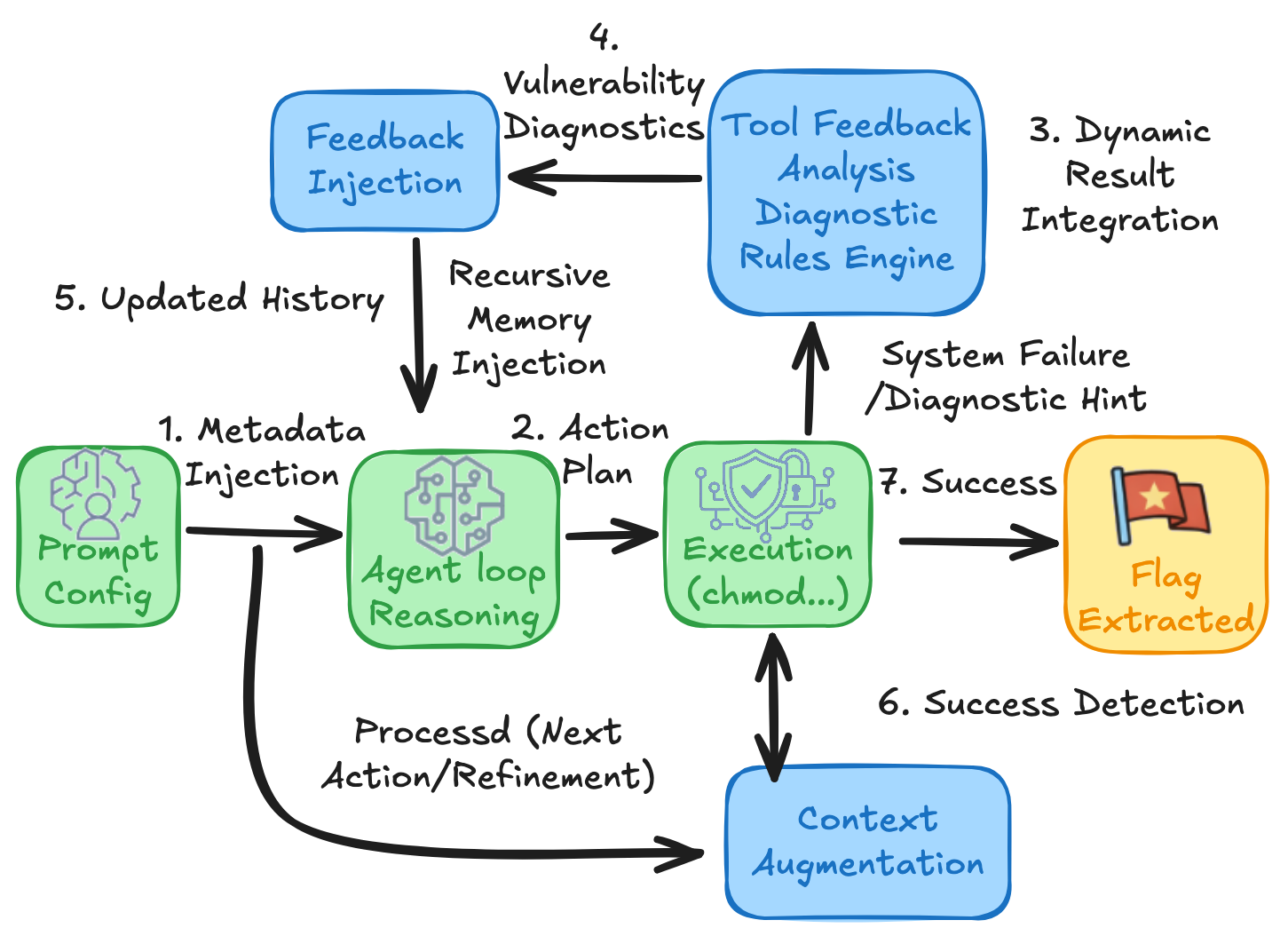}
    \caption{Single agent grounded loop architecture.}
    \label{fig:In_class_agent_a}
  \end{subfigure}

  \vspace{0.6em}

  \begin{subfigure}[t]{\linewidth}
    \centering
    \includegraphics[width=0.9\linewidth]{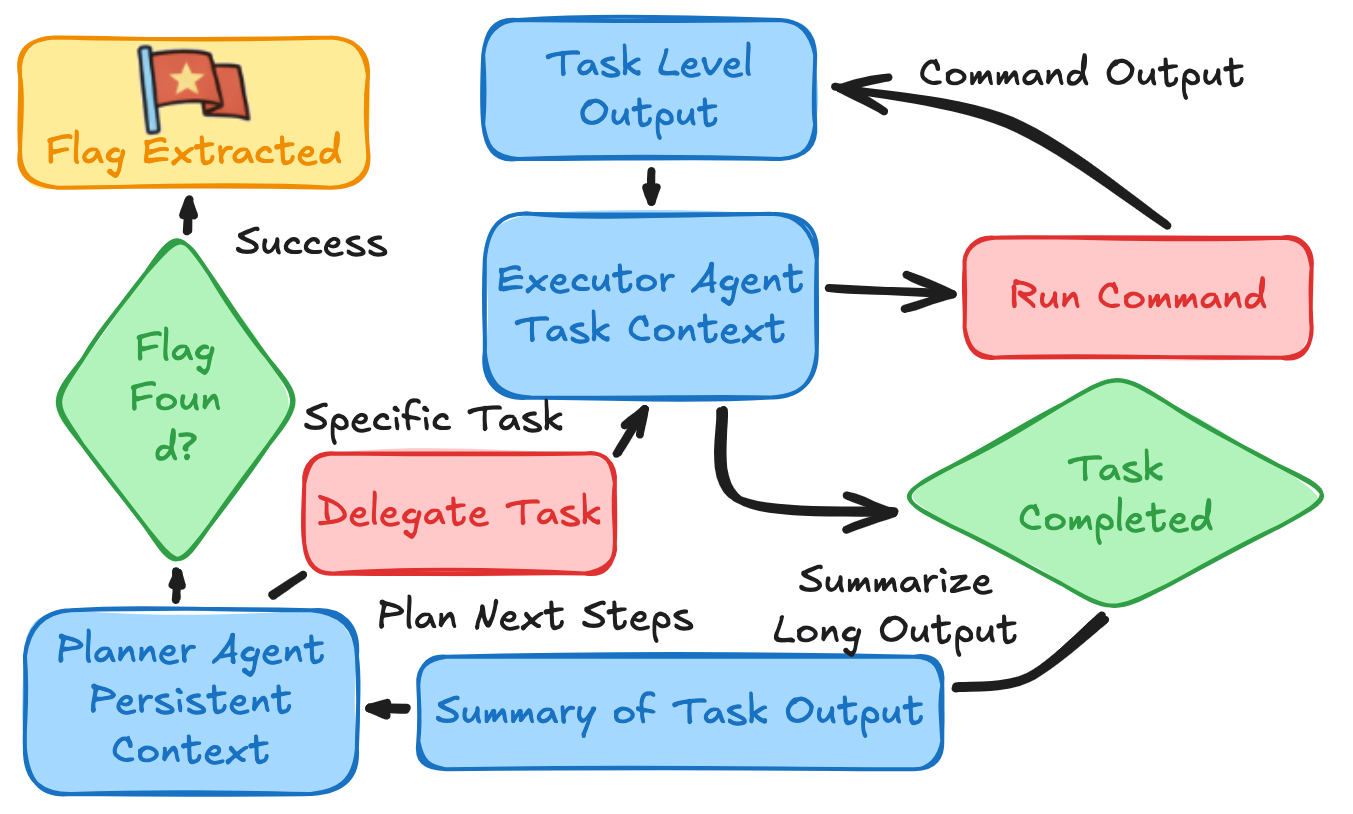}
    \caption{Planner-Executor multi-agent architecture.}
    \label{fig:In_class_agent_b}
  \end{subfigure}

  \caption{In-class Track agent architecture workflows.}
  \label{fig:In_class_agent}
\end{figure}

In Table~\ref{tab:inclass_arch_and_addons}, we structure Case A2 using a two layer taxonomy that separates agent architecture from techniques. In the single-agent grounded loop in Figure~\ref{fig:In_class_agent_a}, a single controller agent receives challenge metadata and configuration, proposes an action plan, calls local tools such as shell, decompiler, and debugger, and  integrates tool outputs back into its context to refine subsequent actions. This architecture tightly couples planning, execution, and interpretation of tool feedback in a single agent, making it straightforward to implement and broadly applicable across challenges. In the planner-executor multi-agent architecture shown in Figure~\ref{fig:In_class_agent_b}, responsibilities are separated: a planner maintains persistent context and decomposes the task into subtasks, while an executor runs commands and produces task outputs. A summarization step feeds results back to the planner, closing a plan $\rightarrow$ delegate $\rightarrow$ execute $\rightarrow$ summarize $\rightarrow$ re-plan loop. This separation reduces context overload and keeps the strategy on track. But it entails design overhead that may explain its low adoption.

\section{Reflections and Future Directions}

This study reflects an initial step toward scalable, learning-oriented LLM-assisted CTF competitions, with opportunities for gradual improvement in future iterations. First, the level of autonomy is likely correlated with prior security experience, engineering capacity, and familiarity with LLM-based systems, introducing potential selection bias. Therefore, higher completion rates among more autonomous teams should not be interpreted as a causal effect of autonomy alone. Second, team counts are imbalanced across autonomy levels and tracks, which limits statistical comparability and makes some patterns sensitive to a few high-performing submissions. Third, our evidence-based analysis improves traceability and integrity, and it also underscores the need for clearer standards for documenting reasoning, tool use, and intermediate validation. However, trace completeness is not equivalent to solution quality or learning outcomes.

Future work will connect trace-derived behaviors to direct learning measurements through pre- and post-assessments. We plan to use controlled within team comparisons between human-in-the-loop and agent runs to separate autonomy effects from prior expertise and engineering investment. We also plan to refine category-specific evaluation criteria for challenges that rely heavily on interpretation or unstructured evidence, and to strengthen reproducibility guidelines by specifying minimum requirements for trace completeness and reporting. More broadly, the track structure, evidence requirements, and autonomy taxonomy are designed to be portable to other AI-assisted cybersecurity competitions and may also inform broader educational assessments beyond CTFs.

\section{Conclusion}

We present a scalable and learning-oriented design for running LLM-assisted CTFs in a way that remains fair for participants with diverse backgrounds. We structure the CTFs into tracks that align with the teams' experience, and we introduce an explicit autonomy taxonomy that separates human-guided usage from autonomous agent workflows. To make results auditable and comparable, we require traceability through complete conversation logs, agent execution trajectories, and released agent code, allowing both outcomes and processes to be examined. Using CTF data collected across regions, we observe that teams operating at higher autonomy levels  achieve higher CTF completion rates, and this pattern is most visible on challenges where rapid iteration, tool interaction, and repeated verification are central to progress. The track-based structure, autonomy taxonomy, and traceability requirements also advance educational goals by emphasizing explainable problem solving, intermediate validation, and reflective human–AI collaboration rather than final outcomes alone. More broadly, the framework and resulting dataset provide a practical foundation for the community to study how autonomy shapes performance and problem solving, design fair and learning-oriented LLM-assisted cybersecurity competitions for diverse participants, and refine prompt engineering and agent architecture choices for effective CTF solving.

\bibliographystyle{IEEEtran}
\bibliography{refs}

\end{document}